\documentclass[12pt,fleqn]{article}

\usepackage{amsmath,amssymb,amsthm}

\begin{document}
\allowdisplaybreaks

\begin{flushleft}
\Large \bf  On supersymmetries in nonrelativistic quantum
mechanics
\end{flushleft}

\begin{flushleft} \bf
J. Beckers~$^\dag$, N. Debergh~$^\dag$ and A.G. Nikitin~$^\ddag$
\end{flushleft}

\noindent
$\dag$~Theoretical and Mathematical Physics, Institute of Physics,\\
$\phantom{\dag}$~B.5, University of Liege, B-4000 Liege I, Belgium\\[2mm]
$\ddag$~Institute of Mathematics of NAS of Ukraine,\\
$\phantom{\ddag}$~3 Tereshchenkivska Str., Kyiv 4, 01601 Ukraine

\begin{abstract}
One-dimensional nonrelativistic systems are studied when time-independent
potential interactions are involved. Their
supersymmetries are determined and their closed subsets generating
kinematical invariance Lie superalgebras are pointed out. The
study of even supersymmetries is particularly enlightened through
the already known symmetries of the corresponding Schr\"odinger
equation. Three tables collect the even, odd, and total
supersymmetries as well as the invariance (super)algebras.
\end{abstract}

\renewcommand{\theequation}{\arabic{section}.\arabic{equation}}

\section{Introduction}

In the seventies, systematic studies of {\it symmetries}
in nonrelativistic quantum mecha\-nics (NRQM) had already been
realized when the Schr\"odinger equation involves
time-independent potential interactions \cite{1,2,3,4}. Some of them have
considered one-dimensional~\cite{2,4}, three-dimensional~\cite{1},
 or $n$-dimensional~\cite{3} (space) systems.

Recently~\cite{5} we have added new information
by visiting higher-order symmetries of such one-dimensional
Schrodinger equations and by considering the third-order context as an effective case,
(Remember that the usual well-known symmetries of the
Schr\"odinger equation correspond to first- and second-order
operators.)

Since the event of supersymmetry in theoretical particle physics~\cite{6}
 and its implications~\cite{7} in supersymmetric
(nonrelativistic) quantum mechanics (SSQM),
 we are dealing with supersymmetric wave equations, including
time-independent superpotentials [hereafter denoted $W(x)$
in one-dimensional problems]. Such supersymmetric
equations are usually put in the following matrix form:
\begin{gather}
i \partial_t \chi(x,t)=\left[ -\frac{1}{2} \partial^2_x +
\frac{1}{2} W^2(x)+\frac{1}{2}W'(x)\sigma_3\right]\chi(x,t),
\end{gather}
where $\partial_t=\partial/\partial t$,
the prime refers to the space derivative $\partial_x \equiv
\partial/\partial x$, $\sigma_3$ is the Pauli $2 \times 2$ matrix diag$(1,-1)$, and
correlatively $\chi(x,t)$ is a two-component wave function.
For simplicity we consider systems with unit
masses and choose $\hbar=1$ in this quantum context.

In Hamiltonian form, Eq.~(1.1) contains as expected, the bosonic
part $H_b$ and the fermionic part $H_f$ given by
\begin{gather}
H_b=-\frac{1}{2} \partial^2_x+\frac{1}{2} W^2(x), \quad H_f=\frac{1}{2}W'(x)\sigma_3, \quad
H_{SS}=H_b+H_f,
\end{gather}
showing that superpotentials $W(x)$ are simply related to usual
potentials $U(x)$ by
\begin{gather}
U(x)=\frac{1}{2}W^2(x).
\end{gather}
Such supersymmetric equations (1.1) and characteristics (1.2)
have already suggested new methods and results~\cite{8,9}
by considering even and odd symmetries in connection with the so-called Lie
extended method developed in quantum physics~\cite{10}.

As a particular context that has been recently visited~\cite{8},
let us mention the case of the one-dimensional
supersymmetric harmonic oscillator corresponding to
\begin{gather}
 W(x)=\omega x\ \Leftrightarrow \ U(x)=\frac{1}{2} \omega^2 x^2.
\end{gather}

Here we want to study systematically
the supersymmetries as well as their superstructures subtended by the
SSQM-Schr\"odinger equation (1.1) for arbitrary
superpotentials $W(x)$. Such a program will appear in the
following sections where we will get a complete classification of all solvable (admissible)
interactions admitting nontrivial supersymmetries.

The contents of this paper are distributed as follows.
In Section~2 we come back on the Boyer results~\cite{3},
but for one-dimensional systems. We present the corresponding study in a particularly
convenient way for our purpose in the supersymmetric context developed in Section~3.
There we first
determine the even supersymmetries (Section~3.1), then second the odd ones (Section~3.2),
and finally we
superpose both contexts (Section 3.3) and get only four priviledged
classes of superpotentials. All the
results are quoted in three respective tables, where we mention the
largest invariance Lie (super)algebras
generated by the corresponding operators.
Remember that these extended (super)symmetries do not
necessarily close. Section 4 finally contains some comments on
isomorphic structures and some conclusions.

\setcounter{equation}{0}

\section{Going back on symmetries in NRQM}

In fact, Boyer~\cite{3} has already solved the problem
of determining all the symmetries of the Schr\"odinger
equation in n dimensions. Applied to the simplest one-dimensional context, this problem
consists in the resolution of the following invariance condition:
\begin{gather}
[\Delta, Q]= i\lambda \Delta, \quad \lambda \equiv \lambda(x,t).
\end{gather}
Here the Schr\"odinger equation defines the $\Delta$ operator on the form
\begin{gather}
\Delta \psi(x,t) \equiv \left(-i \partial_t-\frac{1}{2} \partial^2_x+U(x)\right)\psi(x,t),
\end{gather}
while we are searching for symmetry operators $Q$ of (at most)
second order with respect to space derivatives, i.e.,
\begin{gather}
Q \equiv ia(x,t) \partial_t+ib(x,t)\partial_x+ic(x,t),
\end{gather}
$a$, $b$, and $c$ being arbitrary functions.
Let us mention that higher-order symmetry operators have
recently been considered elsewhere in the same context~\cite{5}.

The above problem leads to a set of partial differential equations
giving rise to $x$-independent functions
$a_0$, $b_0$, and $c_0$
and to the following first-order partial differential equation on $U(x)$:
\begin{gather}
\left(\frac{1}{2}\dot a_0 x+b_0\right)U'(x)+\dot a_0U(x)
=-\frac{1}{4}\mathop{a}\limits^{\ldots}\!{}_0 x^2-\ddot b_0x-i
 \dot c_0+\frac{i}{4}\ddot a_0,
 \end{gather}
where the overdots evidently refer to time derivatives. The
general solution of Eq.~(2.4) appears as the sum of the general
solution $U_0(x)$ of the homogeneous equation and of a particular
solution $U_1(x)$ of the inhomogeneous equation. Due to the at most
quadratic dependence on $x$ in the inhomogeneous part of Eq.
(2.4), Boyer has proposed to search for the general solution~\cite{3}
\begin{gather}
U(x)=U_0(x)+U_1(x),
\end{gather}
with
\begin{gather}
U_1(x)=\frac{1}{2}\alpha x^2+\beta x+\gamma,
\quad \alpha, \beta, \gamma =\mbox{arbitrary constants},
\end{gather}
and to distinguish three cases according to $\alpha= 0$,
$\alpha=\omega^2 > 0$, or $\alpha=-\omega^2 < 0$.

If the whole discussion takes place in Boyer's work~\cite{3}, let us
notice here the following results and properties in order to
exploit them in the supersymmetric context.

In the free case, i.e., $U=0$ $(U_0 = 0 = U_1)$, as already obtained by Niederer~\cite{1},
we get at most six symmetries whose
commutation relations lead to the largest invariance algebra seen as a semidirect
sum $(\Box)$ of the so-called
``conformal'' algebra so$(2,1)$ and the Heisenberg algebra $h(2)$.

In the interacting case but when $U_0=0$,
we want to point out a new result, i.e., the potential
\begin{gather}
U(x)=U_1(x)
\end{gather}
leads to isomorphic structures so$(2,1) \Box h(2)$ a
s in the free case $(\alpha=\beta=\gamma=0)$
whatever are the constants $\alpha$, $\beta$, and $\gamma$.
As particular cases, the harmonic oscillator
corresponds to $U_0=0$, $\alpha=\omega^2$,
$\beta=\gamma=0$ [see Eqs. (1.3) and (2.7)] while the linear
potential to $U_0=0$, $\alpha=0$, i.e.,
\begin{gather}
U(x)=\beta x+\gamma.
\end{gather}
Consequently, these cases also admit six symmetries
and there exist changes of variables connecting all the equations
including the potential forms (2.7) for arbitrary $\alpha$, $\beta$ and~$\gamma$,
with the free Schr\"odinger equation. Let us just quote
as an example some formulas corresponding to the
change of variables between the case $\alpha = \omega^2>0$
and the free case $(\alpha=\beta=\gamma= 0)$.
It can be shown that the relations
\begin{gather}
t_1=(1/\omega) \tan^{-1} \omega t_2, \quad
x_1=(1+\omega^2t^2_2)^{-1/2}(x+\beta /\omega^2)-\beta/\omega^2
\end{gather}
correspond to a well-defined change
of variables between the above interacting case (indices 1) and the free
case (indices 2) implying a modification in the
corresponding Schr\"odinger wave functions according to
\begin{gather}
\Psi_1(x_1,t_1)=(1+\omega^2t^2_2)^{1/4} \exp \left[\left(
\frac{i\beta^2}{2\omega^3}-\frac{i\gamma}{\omega}
\right)\tan^{-1}\omega t_2-\frac{i\beta^2}{2\omega^3}\frac{t_2}{1+\omega^2t^2_2}\right]\nonumber\\
\phantom{\Psi_1(x_1,t_1)=}{}\times
\exp \left[-\frac{it_2x}{2(1+\omega^2t^2_2)}(2\beta+\omega^2x) \right] \Psi_2(x_2,t_2).
\end{gather}
As a more particular case included in this comment, i.e., the harmonic oscillator case with
$\alpha = \omega^2$, $\beta= \gamma= 0$, we immediately recover the Niederer result~\cite{1}:
\begin{gather}
\Psi_1(x_1,t_1)=(1+\omega^2t^2_2)^{1/4} \exp
\left[-\frac{it_2\omega^2x^2}{2(1+\omega^2t^2_2)}\right]
\Psi_2(x_2, t_2).
\end{gather}

When nontrivial potentials $U_0(x)\not=0$ are considered, only two cases have to be mentioned:
either $U_0$ is arbitrary, then all the potentials
\begin{gather}
U(x)=U_0+\frac{1}{2}\alpha x^2+\beta x+\gamma
\end{gather}
lead   to at least two  symmetries (the corresponding Hamiltonian
and the identity operator); or $U_0$
takes the nonzero form
\begin{gather}
U_0(x)=\delta/(\mu x+\epsilon)^2, \quad \mu \not =0,
\end{gather}
then all the potentials
\begin{gather}
U(x)=\delta/(\mu x+\epsilon)^2+\frac{1}{2} \alpha x^2+\beta x+ \gamma,\quad
[(\mu,\epsilon)=(\alpha,\beta) \ \mbox{if} \ \alpha \not=0],
\end{gather}
lead to four symmetries generating the direct sum ${\rm so}(2,1) \oplus I$.

Such properties are of special interest for the following
discussion of the supersymmetries of Eq.~(1.1) in particular.

\setcounter{equation}{0}

\section{Going to supersymmetries in SSQM}

Already considered \cite{8,9}
for only one-dimensional supersymmetric harmonic oscillators characterized by
interacting terms given in Eq. (1.4),
the search for the largest number(s) of one-parameter Lie algebras can
be extended to arbitrary (one-dimensional) systems. This asks for considering
the problem (2.1) but with an operator ASS defined by the
supersymmetric equation (1.1) and with symmetry
operators $Q$ containing even $(\vec 0)$ and odd (1) parts~\cite{8} according to
\begin{gather}
Q=Q_0+Q_1,
\end{gather}
refering to the expected graduation in the supercontext.
Consequently, let us decompose our program in three steps: first
to study the even supersymmetries (Section 3.1) by exploiting the
results contained in Section~2 on symmetries; second to develop
the new odd context and to get the corresponding supersymmetries
(Section~3.2); and, third, to superpose the two sets of results
(Section~3.3) in order to obtain the complete classification of
solvable interactions in SSQM.

Let us just notice here that, in connection with Eq. (1.1), we
understand that even and odd considerations are directly
connected with $2\times 2$ Pauli matrices or more correctly with the
Clifford algebra $Cl_2\equiv \{\sigma_0\equiv I_2, \sigma_1, \sigma_2, \sigma_3\}$,
where we easily
distinguish the even matrices $(\sigma_0,\sigma_3)$
and the odd ones $(\sigma_1, \sigma_2)$
the usual fundamental representation~\cite{11}.
Such a remark directly enlightens Sections 3.1 and 3.2.

\subsection{Even supersymmetries in SSQM}

The invariance condition (2.1) is replaced here by
\begin{gather}
[\Delta_{SS}, Q_0]=i\lambda_0\Delta_{SS}, \quad \lambda_0=\lambda_0(x,t),
\end{gather}
where
\begin{gather}
\Delta_{SS}=\left(\!\! \begin{array}{cc}
-i\partial_t-\frac{1}{2}\partial^2_x+\frac{1}{2}W^2(x)+\frac{1}{2}W'(x) &  0 \vspace{2mm}\\
 0  & -i\partial_t-\frac{1}{2}\partial^2_x+\frac{1}{2}W^2(x)-\frac{1}{2}W'(x)\end{array}\!\!
  \right)\!\!
  \end{gather}
and
\begin{gather} Q_{0}=\left( \begin{array}{@{}c@{}c@{}}
\begin{array}{@{}c@{}}
-i(a_0+a_3)\partial_t+i(b_0+b_3)\partial_x\\ {} +i(c_0+c_3)\end{array} &  0 \vspace{2mm}\\
 0  & \begin{array}{@{}c@{}} i(a_0-a_3)\partial_t+i(b_0-b_3)\partial_x\\
 {}+i(c_0-c_3)\end{array} \end{array}
 \right) \end{gather}
due to the explicit forms of the even matrices $\sigma_0$ and $\sigma_3$.
Such a problem is equivalent to a set of two
distinct ones in NRQM, as discussed in Section~2.
Indeed we get the two invariance conditions
\begin{subequations}
\begin{gather}
\left[-i\partial_t-\frac{1}{2}\partial^2_x+V_1(x), i(A_1\partial_t+B_1\partial_x+C_1)\right]=i\Lambda_1
\left(-i \partial_t-\frac{1}{2}\partial^2_x+V_1(x)\right)
\end{gather}
and
\begin{gather}
\left[-i\partial_t-\frac{1}{2}\partial^2_x+V_2(x), i(A_2\partial_t+B_2\partial_x+C_2)\right]=i\Lambda_2
\left(-i \partial_t-\frac{1}{2}\partial^2_x+V_2(x)\right),
\end{gather}
\end{subequations}
where the arbitrary functions $A_i$, $B_i$, $C_i$ and $\Lambda_i$ $(i= 1,2)$ are
evidently simply related to the above $a_0, a_3, \ldots$, while we have defined
\begin{subequations}
\begin{gather}
V_1(x)=\frac{1}{2}W^2(x)+\frac{1}{2}W'(x)
\end{gather}
and
\begin{gather}
V_2(x)=\frac{1}{2}W^2(x)-\frac{1}{2}W'(x),
\end{gather}
\end{subequations}
these functions being nothing else than the potentials associated with superpartners~\cite{12}.

\newpage

\begin{center}
\footnotesize

{\bf Table 1.}\\[2mm]
\begin{tabular}{@{}c@{}c@{}c@{}c@{}c@{}}
\hline
$\begin{array}{@{}c@{}} \mbox{Clas-}\\
\mbox{ses}\end{array}$ &
$\begin{array}{@{}c@{}} \mathcal{E} \ \mbox{Super-} \\
\mbox{symmetries and}\\
\mbox{their Lie algebras}\end{array}$ &
$\begin{array}{c@{}} \mbox{Number of}\\
\mathcal{E} \ \mbox{super-}\\
\mbox{symmetries}\end{array}$
   & $\begin{array}{c@{}} \mbox{Explicit forms of associated}\\
\mbox{superpotentials}\end{array}$  & Characteristics \\
\hline
$1$ & $\begin{array}{@{}c@{}}[{\rm so}(2,1)\Box h(2)] \\
\oplus\,[{\rm so}(2,1)\Box h(2)]\end{array}$& $12$
& $W(x)=ax+b$            & $\begin{array}{@{}c@{}} \mbox{Free case}\\
\mbox{Linear case} \\ \mbox{Harmonic} \\ \mbox{oscillator~\cite{9}} \end{array}$ \\
$2$ & $\begin{array}{@{}c@{}}[{\rm so}(2,1)\Box h(2)]\\
\oplus\, [{\rm so}(2,1)\Box {\rm gl}(1)]\end{array}$ & $10$
& $W(x)=\pm 1/(x+c)$     & Coulomb $(c=0)$ \\
$3$ &                             &      & $W(x)=ax\pm 1/x$      & Calogero $(a=\omega)$ \\
$4$ & $\begin{array}{@{}c@{}}[{\rm so}(2,1)\oplus {\rm gl}(1)]\\
\oplus\, [{\rm so}(2,1)\oplus {\rm gl}(1)]\end{array}$
 & $8$  & $W(x)=ax+c/x$             & $a\not=0$, $c\not=\pm 1$ \\
$5$ &                               &      & $W(x)=(fx+g)/(cx^2+dx+h)$ &
$\begin{array}{@{}c@{}} f\not=0,\ c\not=0, \\
f\not= \pm c, \ c \not=0 \end{array}$ \\
$6$ & $\begin{array}{@{}c@{}}[{\rm so}(2,1)\Box h(2)]\\
\oplus\,[{\rm so}(2)\oplus {\rm gl}(1)]\end{array}$
& $8$ & $W(x)=ax+b+ \frac{ce^{\mp ax^2 \mp 2bx}}{d+cfe^{\mp ax^2 \mp 2bx}dx}$
& $\begin{array}{@{}c@{}} (a,b)\not=(0,0) \\
d \not=0, \ c\not=0\end{array}$ \\
$7$  &&& $W(x)=\frac{c}{dx+f}+\frac{g}{h(dx+f)^{\pm 2c/d}\pm [g/(d\mp 2c)](dx+f)}$ &
$\begin{array}{@{}c@{}} d\not= \pm 2c \\
g \not=0, \ c \not=0\end{array}$ \\
 $8$ & $\begin{array}{@{}c@{}}[{\rm so}(2,1)\oplus {\rm gl}(1)]\\
\oplus\,[{\rm so}(2)\oplus {\rm gl}(1)]\end{array} $ & $6$ & $W(x)=\frac{c}{\pm 2cx+d}+
\frac{f}{(\pm 2 cx+d)(g+(f/2c)\ln |\pm 2cx+d|)}$ & $\begin{array}{@{}c@{}} f \not=0, \
a \not=0,\\
d\not=0\end{array} $\\
$9$  &&& $W(x)=ax+b+\frac{c}{dx+f}+\frac{g(dx+f)^{\mp 2c/d} e^{\mp ax^2\mp 2bx}}
{h\pm gfe^{\mp ax^2\mp 2bx}(dx+f)^{\mp 2c/d}dx}\!$ & $ c \not=0, \pm 1$ \\
$10$  & $\begin{array}{@{}c@{}} [{\rm so}(2)\oplus {\rm gl}(1)]\\
\oplus\,[{\rm so}(2)\oplus {\rm gl}(1)]\end{array}$
 & $4$ & $W(x)\not= \ \mbox{above  forms} $ & $\cdots $ \\
\hline
\end{tabular}
\end{center}

By combining and superposing the potentials (2.6), (2.7), (2.12),
and (2.14) in known NRQM, we immediately deduce the
cases leading to (at most) 12 even supersymmetries
or to 10, 8, 6, and (at least) 4 of them. The (six) associated
invariance Lie algebras are also easily determined (see Table~1): the largest
one is seen as the direct sum $[{\rm so}(2,1)\Box h(2)]\oplus [{\rm so}(2,1)\Box h(2)]$
and the smallest one as $[{\rm so}(2)\oplus {\rm gl}(1)]\oplus [{\rm so}(2)\oplus {\rm gl}(l)]$.
Such direct sums evidently come from
the superposition of both projections by $P_\pm =\frac{1}{2}(\sigma_0 \pm \sigma_3)$
of the ordinary symmetries
obtained in NRQM. Moreover, by exploiting the explicit forms
obtained for $V_1$, and $V_2$ in NRQM, it is possible
through Eqs. (3.6) to get the correspon\-ding superpotentials $W(x)$ entering in Eq.~(1.1).
Let us, for example, treat the case leading to the maximal number of even
supersymmetries, i.e., let us consider
\begin{gather}
V_1(x)=\frac{1}{2}\alpha_1 x^2+\beta_1 x+\gamma_1
\end{gather}
and
\[
V_2(x)=\frac{1}{2}\alpha_2 x^2+\beta_2 x+\gamma_2.
\]
Through Eqs. (3.6) we readily get the linear superpotential
\begin{gather}
W(x)=ax+ b, \quad a,b =  \mbox{const},
\end{gather}
which also corresponds to the free case $(a= b= 0)$ as well as to
the one-dimensional harmonic oscillator $(a =\omega$, $b= 0)$. Here are
associated 12 even Supersymmetries, an old result quoted in the even notations~\cite{9},
\begin{gather}
[(H_B, C_{\pm}, I, P_{\pm})\sigma_0] \quad \mbox{and} \quad
  [(H_B, C_{\pm}, I, P_{\pm}) \sigma_3].
   \end{gather}

The general discussion leads to specific families of supcrpotentials
given in Table~1 simultaneously with their
associated invariance Lie algebras
and their dimensions. Let us point out the Coulomb-like and
Calogero-like forms admitting ten even supersymmetries.

\subsection{Odd supersymmetries in SSQM}

With the operator $\Delta_{SS} \equiv (3.3)$
we have now to exploit the invariance condition
\begin{gather}
[\Delta_{SS}, Q_1]=i\lambda_1\Delta_{SS},
\quad \lambda_1=\lambda_1(x,t),
\end{gather}
where
\begin{gather}
  Q_{1}=\left(\begin{array}{@{}c@{}c@{}}
   0  &  \begin{array}{@{}c@{}} (ia_1+a_2)\partial_t+(ib_1+b_2)\partial_x\\
   {} +ic_1+c_2 \end{array} \\
 \begin{array}{@{}c@{}}(ia_1-a_2)\partial_t+(ib_1-b_2)\partial_x\\
 {}+ic_1-c_2\end{array} & 0  \end{array} \right)
\end{gather}
due to the explicit forms of the odd matrices $\sigma_1$ and $\sigma_2$. Such a
problem leads to a set of two third-order equations, which take the following forms,
where the
functions $\alpha_i$, $\beta_i$, and $\gamma_i$, are the $x$-independent parts of
$a_i$, $b_i$, and $c_i$ $(i=1,2)$, respectively:
\begin{subequations}
\begin{gather}
\frac{i}{8}\alpha_2W^{'''}-\frac{3i}{4}\alpha_2 W^2W'-\frac{i}{4}\ddot \alpha_2 x^2W'
-i \dot \beta_2 x W'
-\frac{1}{4}\dot \alpha_2W'+\gamma_2W'\nonumber\\
\qquad {}-\frac{i}{4}\dot \alpha_1 W^2-\frac{i}{2}\ddot \alpha_2 xW-i\dot \beta_2W
 -\frac{i}{4} \mathop{\alpha}\limits^{\ldots}\!{}_1x^2-i \ddot \beta_1x
 -\frac{1}{4}\ddot \alpha_1+\dot \gamma_1=0
\end{gather}
and
\begin{gather}
-\frac{i}{8}\alpha_1W^{'''}+\frac{3i}{4}\alpha_1 W^2W'+\frac{i}{4}
\ddot \alpha_1 x^2W'+i \dot \beta_1 x W'
+\frac{1}{4}\dot \alpha_1W'-\gamma_1W'\nonumber\\
\qquad{}-\frac{i}{4}\dot \alpha_2 W^2+\frac{i}{2}\ddot \alpha_1 xW+i\dot \beta_1W
 -\frac{i}{4} \mathop{\alpha}\limits^{\ldots}\!{}_2x^2-i \ddot \beta_2 x
 -\frac{1}{4}\ddot \alpha_2+\dot \gamma_2=0.
 \end{gather}
\end{subequations}
These inhomogeneous nonlinear equations simultaneously
admit the particular solution (3.8), so that
we can in this supersymmetric context search for the general solution
\begin{gather}
W(x)=W_0(x)+W_1(x),
\end{gather}
where we accept
\begin{gather}
W_1(x)=ax+b, \quad a,b=\mbox{const},
\end{gather}
and where we have to discuss the two cases $W_0=0$
or $W_0\not =0$. Such a discussion is analogous to the one
developed in NRQM through Eqs. (2.5), (2.7), and (2.12).

Let us (first) consider $W_0(x)=0$, so that
\begin{gather}
W(x)=W_1(x)=ax+ b.
\end{gather}
Here again we can distinguish $a=0$ and $a\not=0$. Both cases lead to
12 (odd) supersymmetries. In particular, if $a=\omega$, $b= 0$,
this context coincides with the supersymmetric
harmonic oscillator already visited and
characterized by this maximal number of odd supersymmetries~\cite{9}.

Then let us (second) take $W_0\not=0$ and insert the value (3.13) in
Eqs. (3.12). For arbitrary $W_0$, it is once again interesting to
distinguish between the two cases $a=0$ and $a\not=0$. Such a
discussion leads to at least two odd super-symmetries, which appear
on the forms
\begin{gather}
Q_1^{(1)}=i\sigma_1\partial_x-\sigma_2W(x) \quad  \mbox{and} \quad
Q_1^{(2)}=i\sigma_2 \partial_x+\sigma_1W(x).
\end{gather}
We also point out that they generate the simplest $N=2$ superalgebra
sqm(2) initially introduced by Witten~\cite{7}. We effectively have
\begin{gather}
\{Q_1^{(1)}, Q^{(1)}_1\}= \{Q_1^{(2)},
Q^{(2)}_1\}=-2\partial^2_x+2W^2(x)+2\sigma_3W'(x) =4H_{SS}= 4i \partial_t, \nonumber\\
 \{Q_1^{(1)}, Q^{(2)}_1\}=0, \quad  [Q_1^{(i)}H_{SS}]=0, \quad i
= 1,2.
\end{gather}

In order to complete our classification of admissible interactions
characterized by odd supersymmetries, we now want to determine the
impact due to each symmetry on the superpotentials $W_0(x)$ left
arbitrary in Eq. (3.13). This can be studied by quoting the general
forms of our arbitrary functions $\alpha_i$, $\beta_i$,  and
$\gamma_i$ (depending only on~$t$) appearing in Eq.~(3.12). When
$a=0$, these functions are obtained in terms of the 12 arbitrary
constants $A_{(0)}$, $B_{(0)}$, $C_{(0)}$, $D_{(0)}$, $E_{(0)}$, $F_{(0)}$,
$G_{(0)}$, $H_{(0)}$, $K_{(0)}$, $L_{(0)}$, $M_{(0)}$, and $N_{(0)}$ (the
subscript refers to this $a=0$ case) as follows:
\begin{gather}
\alpha_1(t)=\frac{1}{2} A_{(0)}t^2+B_{(0)}t+C_{(0)}, \nonumber\\
\alpha_2(t)=\frac{1}{2} D_{(0)}t^2+ E_{(0)}t+ F_{(0)}, \nonumber\\
 \beta_1(t)=- \frac{1}{4}b D_{(0)}t^2+ G_{(0)}t+ H_{(0)},\nonumber\\
 \beta_2(t)= \frac{1}{4}b A_{(0)}t^2+ K_{(0)}t+ L_{(0)},\nonumber\\
\gamma_1(t)=\frac{1}{4}A_{(0)}t\left(1+\frac{3i}{2}b^2t\right)
+\frac{i}{4}B_{(0)}b^2t+iK_{(0)}bt+M_{(0)},\nonumber\\
\gamma_2(t)=\frac{1}{4}D_{(0)}t\left(1+\frac{3i}{2}b^2t\right)
+\frac{i}{4}E_{(0)}b^2t-iG_{(0)}bt+N_{(0)},
\end{gather}
while, when $a\not=0$, they are given by
\begin{gather}
\alpha_1(t)=A \exp
(iat)+B\exp(-iat)+iC \exp(3iat)-iD\exp(-3iat),\nonumber\\
\alpha_2(t)=E \exp (iat)+F\exp(-iat)+C \exp(3iat) + D\exp(-3iat),\nonumber\\
\beta_1(t)=\frac{i}{2} b[A\exp(iat)-B \exp(-iat)]+i[K \exp(2iat)-L
\exp(-2iat)]\nonumber\\
\phantom{\beta_1(t)=}{}
-\frac{3}{2}b[C\exp(3iat)-D\exp(-3iat)]+M,\nonumber\\
\beta_2(t)=\frac{1}{2}b[E \exp(iat)-F
\exp(-iat)]+K\exp(2iat)+L\exp(-2iat)\nonumber\\
\phantom{\beta_2(t)=}{}+\frac{3i}{2}b [C \exp(3iat) - D \exp(-3iat)]+ G,\nonumber\\
\gamma_1(t)=\exp(iat)\left(iP+\frac{1}{4}aE+\frac{1}{4}b^2E+\frac{i}{4}aA+\frac{i}{4}b^2A
\right)\nonumber\\
\phantom{\gamma_1(t)=}{}-\exp(-iat)
\left(iQ-\frac{1}{4}aF+\frac{1}{4}b^2F+\frac{i}{4}aB-\frac{i}{4}b^2B
\right)\nonumber\\
\phantom{\gamma_1(t)=}{}
+2ib[K \exp(2iat)+L\exp(-2iat)]-C
\exp(3iat)\left(\frac{3}{4}a+\frac{9}{4}b^2\right)\nonumber\\
\phantom{\gamma_1(t)=}{}
-D\exp(-3iat)\left(\frac{3}{4}a-\frac{9}{4}b^2\right),\nonumber\\
\gamma_2(t)=P\exp(iat)+Q\exp(-iat)+2b
[K\exp(2iat)-L\exp(-2iat)]\nonumber\\
\phantom{\gamma_2(t)=}{}
+\frac{3i}{4}C\exp(3iat)
(3b^2+a)+\frac{3i}{4}D\exp(-3iat)(3b^2-a),
\end{gather}
showing once again 12 arbitrary constants.

As an example, let us introduce (3.13) and (3.14) in Eqs. (3.12)
when $a=0$, i.e., when $W(x)=W_0(x) + b$. By exploiting the
relations (3.18) when only the constant $A_{(0)}$ is nonzero, we
finally obtain three time-independent conditions on $W_0(x)$, which
are
\begin{gather*}
W^{'''}_0-6W_0(W_0+2b)W'_0=0,
\\
2bxW'_0+W_0(W_0+4b)=0,
\\
x(xW'_0+2W_0)=0.
\end{gather*}
They are easily handled for getting the unique solution $W_0(x)=0$.
This $A_{(0)}$ context leads to the symmetry operator
\begin{gather}
Q_1^{A(0)}=\sigma_1\left(\frac{i}{2}t^2\partial_t+ \frac{i}{2}tx\partial_x+
\frac{1}{4}x+ \frac{1}{4}bt^2W-\frac{1}{8}t^2W^2 +\frac{i}{4}
t-\frac{3}{8}b^2t^2\right)\nonumber\\
\phantom{Q_1^{A(0)}=}{}+
\sigma_2\left(
\frac{i}{4}bt^2\partial_x+\frac{1}{2}btx+\frac{i}{8}t^2W'-\frac{i}{4}t^2W\partial_x-\frac{1}{2}txW\right).
\end{gather}
By collecting all the similar information for the whole set of odd
(super)symmetries associated with the case $a=0$, we finally
determine besides the maximal (12) and minimal~(2) numbers of odd
supersymmetries already obtained that only three intermediate cases
can occur: either $W(x)=\pm 1/x$ admits ten (odd) symmetries; or
$W(x)=\pm c/x$ admits four (odd) symmetries if $c \not=\pm 1$; or
$W(x) = b + W_0(x)$ admits four (odd) symmetries if we take account
of an extradependence in terms of Legendre functions.

When $a\not=0$, the developments are more elaborate but the complete
results can once again be obtained through the resolution of
relatively complicated nonlinear differential equations. Just as an
example with
\begin{gather}
W(x)=W_0(x)+ax+b, \quad a\not=0,
\end{gather}
let us use Eqs. (3.12), (3.13), (3.14) and insert Eq. (3.19) when
only the arbitrary constant $C$ is nonzero. This case is then
characterized by the functions
 \begin{gather}
 \alpha_1(t)=iC \exp(3iat), \quad
\alpha_2(t)=C \exp(3iat),\nonumber\\
\beta_1(t)=-\frac{3}{2}bC \exp(3iat), \quad \beta_2(t)=\frac{3i}{2}bC
\exp(3iat),\nonumber\\
\gamma_1(t)=-\frac{3}{4}aC \exp(3iat)-\frac{9}{4}b^2C \exp(3iat),
\nonumber\\
\gamma_2(t)=\frac{9i}{4}b^2C \exp(3iat)+\frac{3i}{4}aC \exp(3iat).
\end{gather}
Thus we get the following third-order (nonlinear) differential
equation on $W_0(x)$:
\begin{gather}
W^{'''}_0=6W^2_0W'_0-12(ax+b)^2W'_0+12(ax+b)W_0W'_0-36a(ax+b)W_0,
\end{gather}
or, in terms of $W(x)\equiv (3.21)$,
\begin{gather}
W^{'''}=6W^2W'-18(ax+b)^2W'-6aW^2-36a(ax+b)W+ 54a(ax+b^2).
\end{gather}
Such an equation has already been quoted and solved in the
literature~\cite{13}: it corresponds to the derivative of a Painlev\'e
IV-type equation and leads to the solution~\cite{13}
\begin{gather}
W(x)=\epsilon\frac{dP_4}{dx}(0,\delta; x)+\frac{\epsilon}{\mu}
\left(x+\frac{b}{a}\right).
\end{gather}

Now, by collecting all the corresponding information for the whole
set of odd (super) symmetries associated with the case $a \not=0$,
we see that only five intermediate cases can occur: Either $W(x)=\pm
1/(ax+ b) + ax + b$ admits ten (odd) symmetries; or $W(x)=c/(ax+b)+
ax+b$ admits four (odd) symmetries if $c\not=\pm 1$; or $W(x) =
W_0(x)+ax+b$ admits four (odd) symmetries if $W_0$ is a solution of
Eq. (3.23); or
\[
W(x)=\frac{d \exp(\pm ax^2)}{d\mp c \int \exp(\pm ax^2)dx}+ax
\]
admits three (odd) symmetries; or
\[
W(x)=\frac{c \exp(\frac{1}{2}b^2x^2 \mp 2bx)}{d\pm c \int \exp(
\frac{1}{2}b^2x^2\mp 2bx)dx} \mp \frac{1}{2}b^2x+b
\]
admits three (odd) symmetries.

Consequently, we can summarize in Table~2 the specific families of
superpotentials admitting odd supersymmetries.

Already mentioned in the even context, we notice once again that the
free case, the linear case, and the harmonic oscillator case are
also involved within the same class $1'$ and that their 12 odd
supersymmetries can be quoted [in correspondence with (3.9)] in the
forms
\begin{gather} [(H_B, C_{\pm}, I, P_{\pm})\sigma_1] \quad \mbox{and} \quad [(H_B,
C_{\pm}, I, P_{\pm}) \sigma_2]
\end{gather}
according to old notations~\cite{9} and remembering that $\sigma_1$ and
$\sigma_2$ are the two odd matrices of $Cl_2$.

\begin{center}
\footnotesize

{\bf Table 2.}\\[2mm]
\begin{tabular}{@{}c@{}c@{}c@{}c@{}}
\hline
Classes &   $\begin{array}{c@{}} \mbox{Number of}\\
\mathcal{E} \ \mbox{super-} \\ \mbox{symmetries}\end{array}$
   & $\begin{array}{c@{}} \mbox{Explicit forms of} \\
   \mbox{associated superpotentials}\end{array}$  & Characteristics \\
\hline
$1'$ &  $12$        & $W(x)=ax+b$      & $\begin{array}{c@{}} \mbox{Free case}\\
\mbox{Linear case} \\ \mbox{Harmonic oscillator \cite{9}}\end{array}$   \\
$2'$ &  $10$        & $W(x)=c/x$       &  $c=\pm 1$ \\
$3'$ &  $10$        & $W(x)=ax+b+c/(x+b/a)$       &  $a\not=0, \ c=\pm 1$ \\
$4'$ &   $4$            & $W(x)=ax+b+c/(x+b/a)$       &  $a\not=0, \ c\not=\pm 1$ \\
$5'$ &          & $W(x)=c/x$       &  $c\not=\pm 1$ \\
$6'$ &   $4$       & $\begin{array}{c@{}} W(x)=b+W_0(x) \\
\mbox{if} \ W^{'''}_0-6W^2_0W'-12bW_0W'_0-6b^2W'_0=0 \end{array}$   &  $a=0$ \\
$7'$ &          & $\begin{array}{c@{}} W(x)=ax+b+W_0(x)\\
\mbox{if} \ W^{'''}_0-6W^2_0W'-12(ax+b)W_0W'_0\\
{} +12(ax+b)^2W'_0+36a(ax+b)W_0=0\end{array}$
 &  $a\not=0$   \\
$8'$ & $3$ & $W(x)=ax+ \frac{c\exp(\pm ax^2)}{d\mp c \int \exp(\pm
ax^2)dx}$   &
$a\not=0, \ c\not=0$ \\
$9'$  &   & $W(x)=\mp
\frac{1}{2}b^2x+b+\frac{c\exp(\frac{1}{2}b^2x^2\mp 2bx)} {d\pm c
\int \exp (\frac{1}{2}b^2x^2\mp 2bx)dx}$ & $b\not=0, \ c\not=0$ \\
$10'$ &  $2$         & $W(x)=W_0(x)+ax+b, \ W_0\not=\mbox{above \ forms} $     & $\cdots$    \\
 \hline
\end{tabular}
\end{center}

\subsection{Supersymmetries and invariance superalgebras in SSQM}

We can now superpose the results on even and odd supersymmetries
collected in Tables~1 and~2, respectively. If such a superposition
is relatively direct for superpotentials belon\-ging to the classes
$1$, $2$, $3$, $4$ and $1'$, $2'$, $3'$, $4'$ due to their similar forms, it is
evident that the superposition of all other information becomes
relatively tedious and that it is not interesting to insist on the
corresponding properties.

Besides the information on the maximal number $24$ $(=12+12)$ of
supersymmetries associated with the classes 1 and $1'$, as well as
on the minimal number $6$ $(=4+2)$ associated with the classes 10 and
$10'$, we can distinguish what are the superstructures generated by
some subsets of operators. We have determined that only three
(closed) invariance Lie superalgebras can be pointed out. We have
summarized these structures and their associated properties in Table~3.
We evidently recover the expected results~\cite{9} for the
supersymmetric harmonic oscillator but also find new ones for the
Coulomb and the Calogero problems.

We notice that only a few cases are priviledged in this
supersymmetric context when we require invariance su
perstructures.

\setcounter{equation}{0}

\section{Comments and conclusions}

As already noticed (and expected), the free case leads to the
largest number (24) of Supersymmetries and to the largest
(13-dimensional) kinematical invariance superalgebra. These results are
also valid for the linear case and the harmonic oscillator context,
which, both, are isomorphic to the free case as it is true in NRQM.
In fact, let us point out that for those three cases we are dealing
with the superpotential given by (see Table~3):
\begin{gather}
W(x)=ax+b,
\end{gather}
so that the corresponding supersymmetric Hamiltonian (1.2) is
\begin{gather}
H_{SS}=\frac 12 p^2 +\frac 12 (a^2 x^2 +2abx+b^2)+\frac 12 a\sigma_3.
\end{gather}
Then, by using the unitary transformation
\begin{gather}
U=\exp[(i/2)at\sigma_3],
\end{gather}
immediately gel for the operator $\Delta_{SS}\equiv (3.3)$, i.e.,
\begin{gather}
\Delta_{SS}\equiv -i\partial_t+H_{SS},
\end{gather}
that
\begin{gather}
U\Delta_{SS}U^{-1}=\left[-i\partial_t-\frac 12 \partial_x^2 +\frac 12(a^2 x^2+2abx+b^2)\right] I_2.
\end{gather}
This expression corresponds to the NRQM context characterized by the
potential (2.6), but amplified by the identity matrix $I_2$
belonging to the Clifford algebra $Cl_2$. We thus recover the six
symmetries obtained by Niederer~\cite{1} multiplied here (four times) by the
ele\-ments of $Cl_2$ leading to $12\mathcal {E}+12\mathcal{O}$, i.e., to 24
supersymmetries~\cite{9} characterized in Eqs.~(3.9) and (3.26). Among them
only 13 close and lead to the largest kinematical invariance
superalgebra ${\rm osp}(2/2)\Box \,{\rm sh}(2/2)$~\cite{14}. Associated with these
comments are changes of variables easily determined from the
formulas (2.9)--(2.11) and directly connected with other results
\cite{1,15} when, for example, $a=\omega$ and $b=0$.

Let us also insist on the specific interest of the simplest Witten
superalgebra sqm(2) considered in Eqs.~(3.17) and recovered as a
part of the minimal closed superstructure found in Table~3. We
immediately notice that
\begin{gather}
{\rm osp}(2/2) \supset {\rm osp}(2/1) \supset {\rm sqm}(2),
\end{gather}
a physical chain of particular interest.

Besides the general conclusions that can be drawn from the tables
and more particularly from Table~3, let us recall that different
types of superpotentials have already been studied in SSQM (see more
particularly the reviews of D'Hoker et al~\cite{16} and of Lahiri et
al~\cite{17}). All these superpotentials fall into one of the classes 1--4
contained in Table~3, as well as those we have considered as
partner potentials in parasupersymmetric quantum mechanics~\cite{18}.

\subsection*{Acknowledgment}

We want to thank Dr. J. Lombet for interesting information
concerning nonlinear differential equations.

\newpage

\begin{center}
\scriptsize

{\bf Table 3.}\\[2mm]
\begin{tabular}{@{}c@{}c@{}c@{}c@{}c@{}c@{}c@{}c@{}}
\hline
&& \multicolumn{3}{@{}c@{}}{Number of supersymmetries} &
\multicolumn{2}{@{}c@{}}{Invariance Lie superalgebras} & \\
 \hline
& &&&&&&  \\[-3mm]
&  Superpotentials & $\mathcal{E}$  &  $\mathcal{O} $  & $N=\mathcal{E}+\mathcal{O}\; $ &
Dimension ($d$) &
Superstructure  & Characteristics \\
$1$ & $W(x)=ax+b$ & $12$ & $12$ &  $24$ & $d=13$ & ${\rm osp}(2/2)\Box
\,{\rm sh}(2/2)$ & $\begin{array}{@{}c@{}} \mbox{Free case}\\ \mbox{Linear case} \\
\mbox{Harmonic} \\ \mbox{oscillator \cite{9}}\end{array}$ \\
$2$ & $W(x)=c/x \left|\begin{array}{@{\,}l@{}} c=\pm 1, \\ c\not= \pm 1, \ c\not=0
\end{array} \right. $ & $\begin{array}{@{}c@{}} 10 \\ 8 \end{array}$  &
$\begin{array}{@{}c@{}} 10 \\ 4 \end{array}$  &
$\begin{array}{@{}c@{}} 20 \\ 12 \end{array}$  &
$\begin{array}{@{}c@{}} d=7 \\ {} \end{array}$  &
 $\begin{array}{@{}c@{}} [{\rm osp}(2/1)\Box \, {\rm so}(2)]\oplus {\rm gl}(1) \\ {}
 \end{array}$ & $\begin{array}{@{}c@{}} \mbox{Coulomb} \\
\mbox{Calogero}\end{array}$ \\
$3$ & $\; W(x)=ax+c/x \left| \begin{array}{@{\,}l@{}}  c=\pm 1, \\ c\not= \pm 1, \ c\not=0
\end{array} \right. $ &
$\begin{array}{@{}c@{}} 10 \\ 8 \end{array}$  &
$\begin{array}{@{}c@{}} 10 \\ 4 \end{array}$  &
$\begin{array}{@{}c@{}} 20 \\ 12 \end{array}$  &
& &  \\
$4$ & $W(x)\not= \mbox{above \ forms} $ & $8,6, \mbox{or} \ 4\;\;$
& $4,3, \mbox{or} \ 2\;\;$ & $12 \leq N \leq 6$
& $d=5$ & $[{\rm sqm}(2)\Box \,{\rm so}(2)]\oplus {\rm gl}(1)$ &$\cdots$  \\
 \hline
\end{tabular}
\end{center}

\end{document}